\documentclass[nofootinbib,aps,letterpaper,preprintnumbers,superscriptaddress,showkeys,11pt]{revtex4}

\usepackage{amsmath}
\usepackage{eurosym}
\usepackage{amssymb}
\pdfoutput=1
\usepackage{graphicx}
\usepackage{color}
\usepackage{braket}
\usepackage{amsmath}
\usepackage{amssymb}
\usepackage{subfigure}
\usepackage{amsfonts}
\usepackage{diagbox}
\usepackage[usenames,dvipsnames,svgnames]{xcolor}  
\usepackage{hyperref}   
\hypersetup{
 breaklinks=true,
pdfstartview={FitH},    
colorlinks=true,       
linkcolor=magenta,          
citecolor=green,        
filecolor=magenta,      
urlcolor=blue,           
anchorcolor=green,      
linktocpage=true
}

\begin{document}
\date{\today}
\newcommand\be{\begin{equation}}
\newcommand\ee{\end{equation}}
\newcommand\bea{\begin{eqnarray}}
\newcommand\eea{\end{eqnarray}}
\newcommand\bseq{\begin{subequations}} 
\newcommand\eseq{\end{subequations}}
\newcommand\bcas{\begin{cases}}
\newcommand\ecas{\end{cases}}
\newcommand{\p}{\partial}
\newcommand{\f}{\frac}

\title{\Large \bf Superradiance and Stability of Kerr Black Hole Enclosed by Anisotropic Fluid Matter}

\author {\textbf{Mohsen Khodadi}}
\email{m.khodadi@ipm.ir}
\affiliation{School of Astronomy, Institute for Research in Fundamental Sciences (IPM)\\
P. O. Box 19395-5531, Tehran, Iran}

\author {\textbf{Reza Pourkhodabakhshi}}
\email{reza.pk.bakhshi@gmail.com
}
\affiliation{Department of Physics, Shahid Beheshti University, G.C., Evin, Tehran 19839, Iran}


\begin{abstract}
Focusing on the rotating black hole (BH) surrounded by the anisotropic fluid matters; radiation, dust, and dark matter, we study the massive scalar superradiant scattering and the stability in the Kiselev spacetime.  Superradiance behavior is dependent on the intensity parameter of the anisotropic matter $K$ in the Kiselev spacetime. By adopting the manifest of low-frequency and low-mass for the scalar perturbation, we find $K<0$ enhances the superradiance scattering within the broader frequency range, compared to $K=0$ while $K>0$ suppresses within the narrower frequency range.  As a result, the radiation and dark matter around the rotating BH act as amplifier and attenuator for the massive scalar superradiance, respectively. This is while the dust has a twofold role because of admitting both signs of $K$. Through stability analysis in the light of the BH bomb mechanism, we show in the presence of dark matter, the instability regime of standard Kerr BH ($K=0$) gets improved in favor of stabilization while the radiation and dust do not affect it. In other words, by taking the dark matter fluid around BH into account, we obtain a broader regime that allows the massive scalar field dynamic to enjoy superradiant stability.

\end{abstract}
\maketitle

\section{Introduction}
Despite some indirect evidence for black holes (BH) such as gravitational waves and dynamical effects on other bodies; finally, by releasing a direct image of the supermassive BH M87* by ``Event Horizon Telescope''(EHT) on 10 April 2019, the existence of BH as a real object in our universe was confirmed.
In the wake of releasing the outcomes of the EHT team, today, more than ever, we are sure BH indeed is a rotating celestial being \cite{Akiyama:2019fyp}. Although, from the causal viewpoint in the classical level, rotating BH serves as one-way membranes, the rotational energy extraction from it, is possible via two classic
well-known processes \footnote{Recently, outlined a new energy extraction mechanism from spinning BHs, so called magnetic reconnection which seems efficient \cite{Comisso:2020ykg}.}: Penrose particle collision \cite{Penrose:1969pc}, and superradiance wave scattering \cite{Zel:1971,Zel:1972,Starobinsky:1973aij}. The existence of the event horizon (EH), as a one-way viscous membrane, and a negative-energy region called ergoregion which converts the spacelike Killing vector to a timelike one, and vice versa, are vital for the occurrence of these phenomena. The non-conflict of these phenomena with the energy conservation law and BH's area; has made them attractive theoretically.
Concerning the superradiance, as our phenomenon of interest in this paper, a bosonic wave field,\footnote{Among bosonic fields, it is well-know that the gravitational wave has the strongest possible superradiance amplification \cite{TP:1974}.} whether massive or massless, with the wave frequency $\omega$, after scattering of the rotating BH with angular velocity of horizon $\Omega_+$ will be amplified if $\omega<m\Omega_+$ (here, $m$ is an azimuthal wave quantum number respect to rotation's axis). It means the excess energy being withdrawn from the BH's rotational energy. Superradiance, in essence, is a generic amplifying process that potentially can occur for any dissipative system with appropriate boundary conditions \cite{Bekenstein:1998nt,Brito:2015oca}. So, the essential player to occur the superradiance in the bed of BH is the existence of the EH.
This is basically because the ingoing boundary condition makes the BH behave akin to a dissipative system. For this reason, it is expected for the superradiance to occur in static charged BHs \cite{Bekenstein:1973mi,DiMenza:2014vpa}, in addition to rotating BHs \cite{Detweiler:1980uk}, and charged rotating BHs \cite{Hod:2016bas,Benone:2019all}. Employing such a way of extracting energy from the BHs is one of the popular candidates for explaining the powering of jets driven by high energy sources, like the active galactic nuclei (AGN). Taking the framework of alternative theories of gravity to investigate BH superradiance is well-motivated; since it is severely sensitive to the geometries subjected to perturbation \cite{Pani:2011gy}-\cite{Jiang:2021whw}.

The superradiance scattering, due to transferring energy from BH to bosonic wave perturbation, under some conditions is prone to generate instabilities in the background \cite{Witek:2012tr}. By settling a reflecting surface, like a mirror outside BH and or enclosing the BH into a polished cavity, because of superradiant modes close to BH being stuck, and subsequently, growing exponentially between EH and a turning point, thereby, the background via a mechanism known as the ``BH bomb'', turns prone to instability \cite{Cardoso:2004nk}-\cite{Dias:2018zjg}. The terminology of the BH bomb, in essence, comes from the seminal paper \cite{Press:1972zz} because the radiation trapped between EH and mirror, eventually, will reach a point where the mirror will break, and leads to expelling the radiation outward just like a bomb \cite{Hod:2014pza, Hod:2016rqd}. Although, within the mentioned scenario, the reflecting boundary condition is artificially provided by a mirror, it is well-known that the Anti-de-sitter (AdS) spacetime as well as the massive scalar field, are able to play such a role naturally \cite{Furuhashi:2004jk}-\cite{Vieira:2021nha} (see also references therein).
One of the interesting theoretical consequences of the superradiant instabilities is the possibility of the appearance of the novel BH solutions with additional parameters violating the no-hair theorem \cite{Herdeiro:2016tmi}-\cite{Rahmani:2020vvv}. As a phenomenological usage of the superradiant instability, it is utilized to impose constraints on ultralight bosons beyond the standard model  \cite{Brito:2014wla}. Therefore, the superradiant instability lets the BH acts as a natural laboratory for particle detection expected from high energy physics. Higher dimension BHs are one of the favorite natural extensions to address the superradiant instability in the interplay with fundamental physics \cite{Aliev:2008yk}-\cite{Ishibashi:2015rya}.

The fact that the astrophysical BHs are not isolated from matter, but may have been enclosed within a profile of it, conducts us to study the role of matter around the BH on the superradiance and its instability as well. 
Such a study is well-justified, since, in the framework of scalar-tensor gravity, it has been shown the matter around BH, affects the massive scalar superradiance of the rotating BH, impressively \cite{Cardoso:2013opa,Cardoso:2013fwa}.
Although, it is not yet clear what kind of matter dominates the region around the BH, by refer to the Kiselev toy model \cite{Kiselev:2002dx,Toshmatov:2015npp}, we pick up a family of exact solutions in which the rotating BH is surrounded by three types of anisotropic fluid matter; radiation, dust, dark matter with Equation-of-States (EoSs)
$\alpha=1/3,~0$ and $-1/3$, respectively \footnote{Newly, for the charged Kiselev  BH surrounded by these three types of anisotropic fluid matter, the superradiance and instability were analyzed \cite{Cuadros-Melgar:2021sjy}. In this regard, it is worth to mention that BH solution surrounded by anisotropic fluid matter, not only does not come from the Kiselev toy model, but also there is another metric \cite{Cho:2017nhx,Kim:2019hfp} to address which is not our concern in this paper.}. Particularly, the nature of superradiance in our analysis is the massive scalar bosons. The superradiance based on the massive scalar perturbation is well-motivated phenomenologically, since, an ultralight scalar boson, beyond the standard model, is expected to exist which is able to fix some highlight issues in cosmology and particle physics, see \cite{Hui:2016ltb} for more details. 

The outline of the current paper is as follows.
In Sec. \ref{sec:Kerr}, we briefly introduce the Kiselev rotating spacetime metric include the anisotropic fluid matter.
By hiring the semi-analytically procedure in Sec. \ref{sec:swb}
, we derive the superradiance amplification factor corresponding to the massive scalar perturbation scattered off the Kiselev rotating background. 
In Sec. \ref{sec:stability}, by analyzing the effective potential in light of the BH bomb mechanism, we discuss the superradiant stability of the dynamics of the massive scalar fields. Eventually, we end this paper with the presentation of a conclusion in Sec. \ref{sec:DC}.
Throughout this paper, we work with the units $c=1=G$.

\section{Kiselev rotating spacetime metric}\label{sec:Kerr}
By employing the Newman-Janis algorithm \cite{Newman:1965tw} as a conventional manner for producing a rotational spacetime metric from its  spherically symmetric counterpart, along with incorporating the modifications proposed in \cite{Azreg-Ainou:2014pra}, the rotating BH solution in the Boyer-Lindquist coordinates $(t,r,\theta,\varphi)$, takes the following form \cite{Toshmatov:2015npp}
\begin{eqnarray}\label{Rot}
	ds^2 &&= -\left(\frac{\Delta_k-a^2\sin^2\theta}{\Sigma}\right)dt^2
	-2 a  \sin^2 \theta \left(\frac{\Delta_k-(r^2+a^2)}{\Sigma}   \right) dt d\varphi+\frac{\Sigma}{\Delta_k}dr^2+\Sigma d\theta^2+ \\ \nonumber
	&&\sin^2 \theta \left(\frac{(r^2+a^2)^2-a^2 \Delta_k \sin^2\theta}{\Sigma}\right)d\varphi^2~,
\end{eqnarray}
where
\begin{eqnarray}
	\Sigma= r^2+a^2 \cos^2\theta,~~~~~\Delta_k=r^2-2Mr+a^2-Kr^{1-3 \alpha}~.
\end{eqnarray}
Here, $M, a,K$ and $\alpha$ are BH mass, rotation parameter of BH, integration constant representing the intensity of fluid matter differing the above metric from standard Kerr ($K=0$),  and EoS representing the type of dominated matter \footnote{In the real universe, these three types of matter may be mixed. However, for simplicity, we assume that the EoS regarded here represents just the dominant effect of different species forms around the BH.} distribution around BH, respectively. The parameter $K$, in essence, controls the density of the fluid enclosing the BH. In this paper, we limit ourselves to three types of fluid matter around BH: 
radiation ($\alpha=1/3$), dust ($\alpha=0$) and  dark matter ($\alpha=-1/3$). Concerning the last case, it is worth mentioning that dark matter, in general, is modeled in different formats which here, it is considered as a
fluid matter \cite{Rahaman:2010xs,Xu:2018mkl}.
The interesting property of the Kiselev metric is that by regarding any of these 
fluid matters, it can recover some of the well-known classical solutions. By setting $\alpha=1/3$, it effectively can play the role of a Kerr-Newman BH; while for the case of $\alpha=0$, we deal with a Kerr BH which its mass got shifted. Interestingly, we see that the case of $\alpha=-1/3$ is very similar to the Kerr metric surrounded by a cloud of cosmic strings instead of point particles \cite{Letelier:1979ej}.
Given these three types of matter, the metric function $\Delta_k$, has two real roots $r_\pm$ where the largest $r_+$ denotes the location of EH. Regarding the metric (\ref{Rot}), we must pay attention to two points. First of all, despite a frequent mistake in literature, the metric (\ref{Rot}) does not address a perfect fluid metric and indeed it represents
some type of anisotropic fluid matter. Second, the positivity of the energy density of the surrounding fluid matter of BH $\rho \geq 0$, as a consequence of the weak energy condition, leads to imposing the constraint $\alpha\, K \leq 0$ on the two additional parameters of the metric (\ref{Rot}).
To investigate these two points in detail it is recommended that refer to Visser's critical paper \cite{Visser:2019brz}. Throughout this paper, we will call the metric (\ref{Rot}) the Kiselev rotating BH include the anisotropic fluid matter.

\section{Superradiance Scattering of Scalar Wave by Kiselev rotating BH Surrounded by anisotropic matter}\label{sec:swb}
In this section, we intend to study the impact of the additional parameters $K$ and $\alpha$, arising from the anisotropic fluid matter on the BH energy extraction by superradiance scattering. Specifically, we carry out this by means of calculating the superradiance amplification factor for a massive scalar field scattered from the Kiselev rotating BH (\ref{Rot}). Worth to note that throughout the paper as usual, we restrict the analysis to the leading order of perturbation and ignore the effect of backreaction of the scalar field on the geometry of background spacetime \cite{Brito:2015oca}.  Indeed, we imagine during the scattering of the scalar field, the spacetime geometry remains fixed without any response to it. In the following, we address the superradiance scattering in light of both the analytical and numerical treatments.
\subsection{Equation of motion}
The governing scalar wave equation, for the scalar field $\Phi$ with the mass $\mu_s$, is a Klein-Gordon (KG) equation as
\begin{equation}\label{eq:KGEq}
	\left(\nabla_\alpha \nabla^\alpha +\mu_s^2\right)\, \Phi(t,r,\theta,\phi)=\bigg(\frac{1}{\sqrt{-g}}\partial_\alpha\big(\sqrt{-g}g^{\alpha\beta}\partial_\beta\big)+\mu_s^2\bigg)\Phi(t,r,\theta,\phi)=0~.
\end{equation}
As usual, in the Boyer-Lindquist coordinates $(t,r,\theta,\phi)$, using the following ansatz
\begin{equation}\label{eq:an}
	\Phi(t,r,\theta,\phi)
	= R_{\omega lm}(r)\; S_{\omega lm}(\theta)\; e^{-i\omega t}\; e^{+i m\phi }~,~~~l\geq0,~~~-l\leq m\leq l,~~~\omega>0,
\end{equation} one can decompose the scalar field function in terms of 
the radial function $R_{\omega lm}(r)$, and the spherical wave function $ S_{\omega lm}(\theta)$. Here, the subscripts $l,~m$, and $\omega$ denote the angular quantum number, the azimuthal wave number and the positive frequency of the scattering scalar field measured by a distant observer. By taking the metric (\ref{Rot}) into the differential equation \eqref{eq:KGEq} as well as importing the ansatz \eqref{eq:an}, it yields two ordinary radial and angular differential equations, as follows
\begin{subequations}
	\begin{align}
		&\frac{d}{d r}\, \left(\Delta_k\, \frac{d R_{\omega lm}(r)}{d r}\right)
		+\left(
		\frac{{\left(\left(r^2 +a^2\right)\omega -a\, m\right)}^2}{\Delta_k}
		-\big(\mu_s^2 r^2 +l(l+1)+a^2\omega^2-2m a\omega\big)\right) R_{\omega lm}(r)= 0~, \label{eq:ODE_rad}\\
		\begin{split}
			&\sin\theta\, \frac{d}{d \theta}\,
			\left(\sin\theta\; \frac{d S_{\omega lm}(\theta)}{d \theta}\right)
			+\bigg(l(l+1)\sin^2\!\theta-\bigg({(a\, \omega\, \sin^2\theta -m)}^2
			+a^2\, \mu_s^2\,\sin^2\theta\cos^2\!\theta \bigg)\bigg) S_{\omega lm}(\theta) = 0~. \label{eq:ODE_ang}
		\end{split}
	\end{align}
\end{subequations}
Within the direction of our aim, we skip the angular equation \eqref{eq:ODE_ang} and instead, we concentrate on the radial equation \eqref{eq:ODE_rad}.
Applying a ``tortoise'' coordinate $r_*$ \footnote{The coordinate $r_*$ maps the range $r\in [r_+,\infty)$ to the whole real axis.} as $\frac{dr_*}{dr}\equiv~ \frac{r^2 +a^2}{\Delta_k}$ ($r_* \rightarrow -\infty$ at EH and $r_* \rightarrow\infty$ at infinity),  plus offering a new radial function $\mathcal{F}_{\omega lm}(r_*)= \sqrt{r^2+a^2}\, R_{\omega lm}(r)$, thereby,
one can re-express the radial equation \eqref{eq:ODE_rad} in the form of a Schr\"odinger-like deferential equation as below
\begin{equation} \label{eq:ODE_rad_Tort}
	\frac{d^2\mathcal{F}_{\omega lm}(r_*)}{dr_*^2}\,
	+U_{\omega lm}(r)\, \mathcal{F}_{\omega lm}(r_*)
	= 0~,
\end{equation}
where
\begin{equation} \label{eq:PotT}
	U_{\omega lm}(r)
	= {\left(\omega -m\Omega\right)}^2
	-\frac{\Delta_k \big(l(l+1)+a^2\omega^2-2ma\omega+\mu_s^2 r^2\big)}{(r^2+a^2)^2}\,  \,
	-\frac{\Delta_k(3r^2-4Mr+a^2)}{(r^2+a^2)^3}+\frac{3r^2\Delta_k^2}{(r^2+a^2)^4}
	~,
\end{equation} is the scattering potential. Here, $\Omega=\frac{ a}{r^2 +a^2}$ is the angular velocity of rotating BH. Due to the vital role of the boundary conditions in the study of scattering process, we need to construct sets of basic modes for the massive scalar field on the EH, as well as the spatial infinity. In this regard, the asymptotically radial solution of \eqref{eq:ODE_rad_Tort} reads as
\begin{equation} \label{eq:fulls}
	\begin{split}
		R_{\omega lm}(r)\to
		\begin{cases}
			\mathcal{I}^{+}~\frac{\,
				e^{-ik_{+}\, r_* }}{\sqrt{r_{+}^2 +a^2}}
			& \text{for $r \to r_{+} ~~(r_* \rightarrow -\infty)$}\\
			\mathcal{I}^{\infty}~
			\frac{e^{-ik_{\infty}\, r_*}}{r}\,
			+ \mathcal{R}^{\infty}~\frac{e^{ik_{\infty}\, r_*}}{r}\,
			& \text{for $r \to \infty~~ (r_* \rightarrow \infty)$}
		\end{cases}
	\end{split}
\end{equation} where
\begin{subequations}
	\begin{align}
		&	\lim_{r\rightarrow r_{+}} U_{\omega lm}(r)
		= {\left(\omega -m\,\Omega_+\right)}^2
		\equiv k_{+}^2~,~~~\Omega_+=\frac{ a}{r_+^2 +a^2}\\
		\begin{split}
			&	\lim_{r\to\infty} U_{\omega lm}(r)
			= \omega^2 - \lim_{r\to\infty}\, \frac{\mu_s^2 r^2\, \Delta_k}%
			{{\left(r^2 +a^2\right)}^2}
			= \omega^2 -\mu_s^2\equiv k_\infty^2,
		\end{split}
	\end{align}
\end{subequations} 
Here, $\mathcal{I}$ and $\mathcal{R}$, represents respectively the incident and reflected parts of the scalar wave at the EH (``+'') or at spatial infinity (``$\infty$'')).  Now, by equalizing  the Wronskian of the regions neighbor the EH, $W_{+}=\big(R_{\omega lm}^{+}\frac{d R_{\omega lm}^{*~+}}{dr_*}
-R_{\omega lm}^{*~+}\frac{d R_{\omega lm}^{+}}{dr_*}\big)$
and those corresponding the regions at infinity $W_{\infty}=\big(R_{\omega lm}^{\infty }\frac{d R_{\omega lm}^{*~\infty }}{dr_*}-R_{\omega lm}^{*~\infty }
\frac{d R_{\omega lm}^{\infty }}{dr_*}\big)$, we finally reach to the following equality between the incident and reflected parts of the amplitude
\begin{eqnarray}\label{flat}
	|\mathcal{R}^{\infty}|^2=|\mathcal{I}^{\infty}|^2-\frac{k_{+}}
	{k_\infty}|\mathcal{I}^{+}|^2\,.
\end{eqnarray} 
As a fascinating feature in the equality aforementioned  can say that it is independent of the scattering potential's details $U_{\omega lm}(r)$ in the Schr\"odinger-like differential equation \eqref{eq:ODE_rad_Tort}. Not difficult to show that the scalar wave is superradiantly amplified, $\frac{|\mathcal{R}^{\infty}|^2}{|\mathcal{I}^{\infty}|^2}>1$, if $\frac{k_{+}}{k_\infty}<0$ i.e. $\omega<m\Omega_{+}$.

\subsection{Superradiant amplification of scalar wave scattering }

Here, we intend to compute the ``amplification factor'' $Z_{lm}\equiv \frac{|\mathcal{R}^{\infty}|^2}{|\mathcal{I}^{\infty}|^2}-1$, of a massive scalar wave scattered off the Kiselev rotating BH enclosed by the anisotropic 
fluid matter.  That way, we are able to track the
effect of additional parameters $k$ and $\alpha$ in the metric (\ref{Rot}) on the amplification factor $Z_{lm}$. This dimensionless quantity is a criterion for occurring BH superradiance if $Z_{lm}>0$.
For this purpose, we need to solve the radial equation \eqref{eq:ODE_rad}. Although, it is not exactly solvable analytically, one can take advantage of a semi-analytically method so-called ``analytical asymptotic matching'' \footnote{Historically, this method comes from the seminal work of Starobinsky in the early 80's \cite{Starobinsky:1973aij}.} (AAM). By hiring the AAM technique, we should assume that the parameters involved in the composed system of BH and scalar perturbation, satisfy the conditions
$a\omega\ll1$ i.e. $\mathcal{O}(a\omega)$
and $M\omega\ll1$ i.e. $\mathcal{O}(\mu_s r_+)$ \cite{Detweiler:1980uk}. The former restricts us to the low-frequency regime of the perturbation, and the latter demands that the BH's size is smaller than  the Compton wavelength associated with the scalar perturbation. In this regard, we actually are dealing with two distinct  zones: \emph{``near-region''} and \emph{``far-region''} which respectively, address around EH ($r-r_{+}\ll \omega^{-1}$) and away from the EH ($r-r_{+}\gg M$). Given that the matching is possible only if the relevant expansions have an overlapping domain, thereby, the exact solutions derived for the two above asymptotic regions are matched inside an overlapping region where $M\ll r-r_{+}\ll \omega^{-1}$. 
In what follows, by utilizing the AAM technique discussed above,
we will obtain a semi-analytical solution for the radial equation \eqref{eq:ODE_rad}.
\subsubsection{\textbf{Near-region solution}}
By offering the change of variable $x=\frac{r-r_+}{r_+-r_-}$
along with definition $\triangle_k \frac{d}{dr}=(r_{+}-r_{-})x(x+1) \frac{d}{dx}$ and also using approximation $a\omega\ll1$, the equation \eqref{eq:ODE_rad} rewrites as 
\begin{align}
	\label{eq:nr}
	&x^{2}(x+1)^2\frac{\mathrm{d}^2F_{\omega lm}(x)}{\mathrm{d}x^2}+x(x+1)(2x+1)
	\frac{\mathrm{d}F_{\omega lm}(x)}{\mathrm{d}x}  \nonumber\\
	&+\left(A x^4+B^2-l(l+1) x(x+1)-\frac{\mu_s^2 A^2}{\omega^2}x^3(x+1)
	-\mu_s^2 r_{+}^2 x(x+1)-\frac{2\mu_s^2 r_{+}^2 A}{\omega} x^2(x+1)\right) F_{\omega lm}(x)=0\,,
\end{align} where 
\begin{eqnarray}\label{const}
	A=(r_{+}-r_{-})\omega,~~~ \mbox{and}~~~ B=\frac{(\omega-m\Omega_+)}{r_{+}-r_{-}}r_{+}^2.
\end{eqnarray}
Because of the two approximations relevant to the regions around th EH i.e. $Ax\ll1$ and $\mu^2r_{+}^2\ll1$, thus the Eq. \eqref{eq:nr}, reduces to
\begin{align}\label{eq:nrr}
	x^{2}(x+1)^2\frac{\mathrm{d}^2F_{\omega lm}(r)}{\mathrm{d}x^2}+x(x+1)(2x+1)\frac{\mathrm{d}F_{\omega lm}(r)}{\mathrm{d}x}
	+\left(B^2-l(l+1) x(x+1)\right)F_{\omega lm}(r)=0\,.
\end{align} 
The general solution of equation \eqref{eq:nrr} satisfying the ingoing boundary
condition, written in terms of ordinary hypergeometric functions $_2F_1(a,b;c;y)$ 
\begin{align}
	F_{\omega lm}(x)=c~\big(\frac{x}{x+1} \big)^{-iB}~_2F_1\left(\frac{1-\sqrt{1+4l(l+1)}}{2},\frac{1+\sqrt{1+4l(l+1)}}{2};1-2iB;-x\right).
\end{align}
As the last step here, it is essential to know that the behavior of the above solution at large $x$ is
\begin{align} \label{eq:solnear}
	F_{\mathrm{near- large~x}}\sim c~\left(\frac{\Gamma\big(\sqrt{1+4l(l+1)}\big)~\Gamma\big(1-2iB\big)}{\Gamma\bigg(\frac{1+\sqrt{1+4l(l+1)}-4iB}{2}
		\bigg)	~\Gamma\bigg(\frac{1+\sqrt{1+4l(l+1)}}{2}\bigg)}
	~x^{\frac{\sqrt{1+4l(l+1)}-1}{2}}+ \right.\nonumber\\\left.
	\frac{\Gamma\big(-\sqrt{1+4l(l+1)}\big)~\Gamma\big(1-2iB\big)}{\Gamma\bigg(\frac{1-\sqrt{1+4l(l+1)}}{2}\bigg)
		~\Gamma\bigg(\frac{1-\sqrt{1+4l(l+1)}-4iB}{2}\bigg)}~x^{-\frac{\sqrt{1+4l(l+1)}+1}{2}} \right)\,.
\end{align} 
\subsubsection{\textbf{Far-region solution}}
In the far-region, after applying approximations as  $x+1\approx x$ and $\mu_s^2r_{+}^2\ll1$, the radial equation \eqref{eq:ODE_rad} reads as
\begin{eqnarray}\label{eq:far}
	\frac{\mathrm{d}^2F_{\omega lm}(x)}{\mathrm{d}x^2}+\frac{2}{x}\frac{\mathrm{d}F_{\omega lm}(x)}{\mathrm{d}x}+
	\left(k^2-\frac{l(l+1)}{x^2}\right)F_{\omega lm}(x)=0\,,~~~k \equiv \frac{A}{\omega}\sqrt{\omega^2-\mu_s^2}~.
\end{eqnarray} 
The general solution of the equation \eqref{eq:far}, would be written as  
\begin{align}\label{sol}
	F_{\omega lm,~\mathrm{far}} = e^{-i k x} \left(d_1~x^{\frac{\sqrt{1+4l(l+1)}-1}{2}}~U\big(\frac{1+\sqrt{1+4l(l+1)}}{2},1+\sqrt{1+4l(l+1)},2ikx\big)+
	\right.\nonumber\\\left.
	d_2~x^{-\frac{\sqrt{1+4l(l+1)}+1}{2}}~U\big(\frac{1-\sqrt{1+4l(l+1)}}{2},1-\sqrt{1+4l(l+1)},2ik x\big)\right)\,,
\end{align} where $U(a,b,y)$ denotes to the first Kummer function.
Here, it is also essential to know the behavior of the above solution at small $x$
\begin{eqnarray}\label{eq:solfar}
	F_{\omega lm,~\mathrm{far-small \,\, x}}\sim  d_1~x^{\frac{\sqrt{1+4l(l+1)}-1}{2}}+
	d_2~x^{-\frac{1+\sqrt{1+4l(l+1)}}{2}}\,.
\end{eqnarray}
\subsubsection{\textbf{Matching of solutions}}
Now, by matching the two asymptotic solutions mentioned above, it yields the scalar wave fluxes at infinity. By means that, we will obtain the expression representing the superradiant amplification factor.
At first step, by facing asymptotic solutions \eqref{eq:solnear} and \eqref{eq:solfar}, we acquire
\begin{eqnarray} \label{eq:d12}
	d_1= c\frac{\Gamma(\sqrt{1+4l(l+1)})~\Gamma(1-2iB)}{\Gamma(\frac{1+\sqrt{1+4l(l+1)}-4iB}{2})~
		\Gamma(\frac{1+\sqrt{1+4l(l+1)}}{2})}\,,~~~
	d_2=c\frac{\Gamma(-\sqrt{1+4l(l+1)})~\Gamma(1-2iB)}{\Gamma(\frac{1-\sqrt{1+4l(l+1)}-4iB}{2})~
		\Gamma\big(\frac{1-\sqrt{1+4l(l+1)}}{2}\big)}\,.
\end{eqnarray}
In this point, we need to connect the coefficients $d_1$ and $d_2$ with the coefficients $\mathcal{I}^{\infty}$ and
$\mathcal{R}^{\infty}$ in the radial solution \eqref{eq:fulls}.
To do that, we expand the far region solution \eqref{sol} around infinity as
\begin{align}\label{eq:Asysoo}
	d_1 \frac{\Gamma(1+\sqrt{1+4l(l+1)})}{\Gamma(\frac{1+\sqrt{1+4l(l+1)}}{2})}k^{-\frac{1+\sqrt{1+4l(l+1)}}{2}}
	\bigg((-2i)^{-\frac{1+\sqrt{1+4l(l+1)}}{2}}\frac{e^{-ik x}}{x}+
	(2i)^{-\frac{1+\sqrt{1+4l(l+1)}}{2}}\frac{e^{ik x}}{x}\bigg)+\\ \nonumber
	d_2 \frac{\Gamma(1-\sqrt{1+4l(l+1)})}{\Gamma(\frac{1-\sqrt{1+4l(l+1)}}{2})}k^{\frac{\sqrt{1+4l(l+1)}-1}{2}}
	\bigg((-2i)^{\frac{\sqrt{1+4l(l+1)}-1}{2}}\frac{e^{-ik x}}{x}+
	(2i)^{\frac{\sqrt{1+4l(l+1)}-1}{2}}\frac{e^{ik x}}{x}\bigg)\, .
\end{align}
By matching the above solution \eqref{eq:Asysoo} with the radial solution \eqref{eq:fulls}, and also by putting
the expressions acquired for the coefficients $d_1$ and $d_2$ in \eqref{eq:d12}, we finally arrive at
\begin{align}\label{eq:Asysooo}
	F_\infty(r)\sim \mathcal{I}^{\infty}~\frac{e^{-i\sqrt{\omega^2-\mu_s^2} r^*}}{r}+\mathcal{R}^{\infty}~
	\frac{e^{i\sqrt{\omega^2-\mu_s^2} r^*}}{r}, \qquad  \mbox{for}~~~r \rightarrow \infty ,
\end{align} where
\begin{eqnarray}\label{eq:A3}
	&&\mathcal{I}^{\infty}=\frac{c~ (-2i)^{-\frac{1+\sqrt{1+4l(l+1)}}{2}}}{\sqrt{\omega^2-\mu_s^2}}.\frac{
		\Gamma(\sqrt{1+4l(l+1)})~\Gamma(1+\sqrt{1+4l(l+1)})}
	{\Gamma\bigg(\frac{1+\sqrt{1+4l(l+1)}-4iB}{2}\bigg)~\bigg(\Gamma(\frac{1+\sqrt{1+4l(l+1)}}{2})\bigg)^2}\times \\ \nonumber
	&&~\Gamma(1-2iB)~k^{\frac{1-\sqrt{1+4l(l+1)}}{2}}+ \frac{c~ (-2 i)^{\frac{\sqrt{1+4l(l+1)}-1}{2}}}{\sqrt{\omega^2-\mu_s^2}}
	~\Gamma(1-2i~ B)~k^{\frac{1+\sqrt{1+4l(l+1)}}{2}}\times\\ \nonumber
	&&\frac{\Gamma(1-\sqrt{1+4l(l+1)})
		~\Gamma(-\sqrt{1+4l(l+1)})}
	{\bigg(\Gamma(\frac{1-\sqrt{1+4l(l+1)}}{2})\bigg)^2 ~\Gamma\big(\frac{1-\sqrt{1+4l(l+1)}-4iB}{2}\big)}\,,
\end{eqnarray}
and
\begin{eqnarray}\label{eq:A4}
	&&\mathcal{R}^{\infty}=\frac{c~(2i)^{-\frac{1+\sqrt{1+4l(l+1)}}{2}}}{\sqrt{\omega^2-\mu_s^2}}.\frac{
		\Gamma(\sqrt{1+4l(l+1)})~\Gamma(1+\sqrt{1+4l(l+1)})}
	{\Gamma\bigg(\frac{1+\sqrt{1+4l(l+1)}-4iB}{2}\bigg)~\bigg(\Gamma(\frac{1+\sqrt{1+4l(l+1)}}{2})\bigg)^2}\times
	\\ \nonumber
	&&\Gamma(1-2iB)~k^{\frac{1-\sqrt{1+4l(l+1)}}{2}}+\frac{c~(2i)^{\frac{\sqrt{1+4l(l+1)}-1}{2}}}{\sqrt{\omega^2-\mu_s^2}}
	\Gamma(1-2iB)~k^{\frac{1+\sqrt{1+4l(l+1)}}{2}}\times\\ \nonumber
	&&\frac{\Gamma(1-\sqrt{1+4l(l+1)})
		~\Gamma(-\sqrt{1+4l(l+1)})}
	{\bigg(\Gamma(\frac{1-\sqrt{1+4l(l+1)}}{2})\bigg)^2 ~\Gamma\bigg(\frac{1-\sqrt{1+4l(l+1)}-4iB}{2}\bigg)}\,.
\end{eqnarray}
Note in deriving \eqref{eq:Asysooo}, the approximations used, are $\frac{1}{x}\sim\frac{A}{\omega}.\frac{1}{r},~~
e^{\pm i k x}\sim e^{\pm i\sqrt{\omega^2-\mu_s^2} r}$.
Now, by having the final form of the incident and reflected coefficients of scalar wave i.e., \eqref{eq:A3} and \eqref{eq:A4}, one can compute the superradiant amplification factor 
\begin{eqnarray}\label{amp}
	Z_{lm}\equiv \frac{|\mathcal{R}^{\infty}|^2}{|\mathcal{I}^{\infty}|^2}-1\,.
\end{eqnarray} 
In the end, let us close this subsection by mentioning the point stating that for the ordinary hypergeometric function $_2F_1(a,b;c;y)$, and the first Kummer function $U(a,b,y)$ in the above calculations, some approximations are used, see \cite{Khodadi:2020cht,Khodadi:2021owg} for further considerations. 
\begin{figure}[!ht]
	\begin{center}
		\includegraphics[scale=0.33]{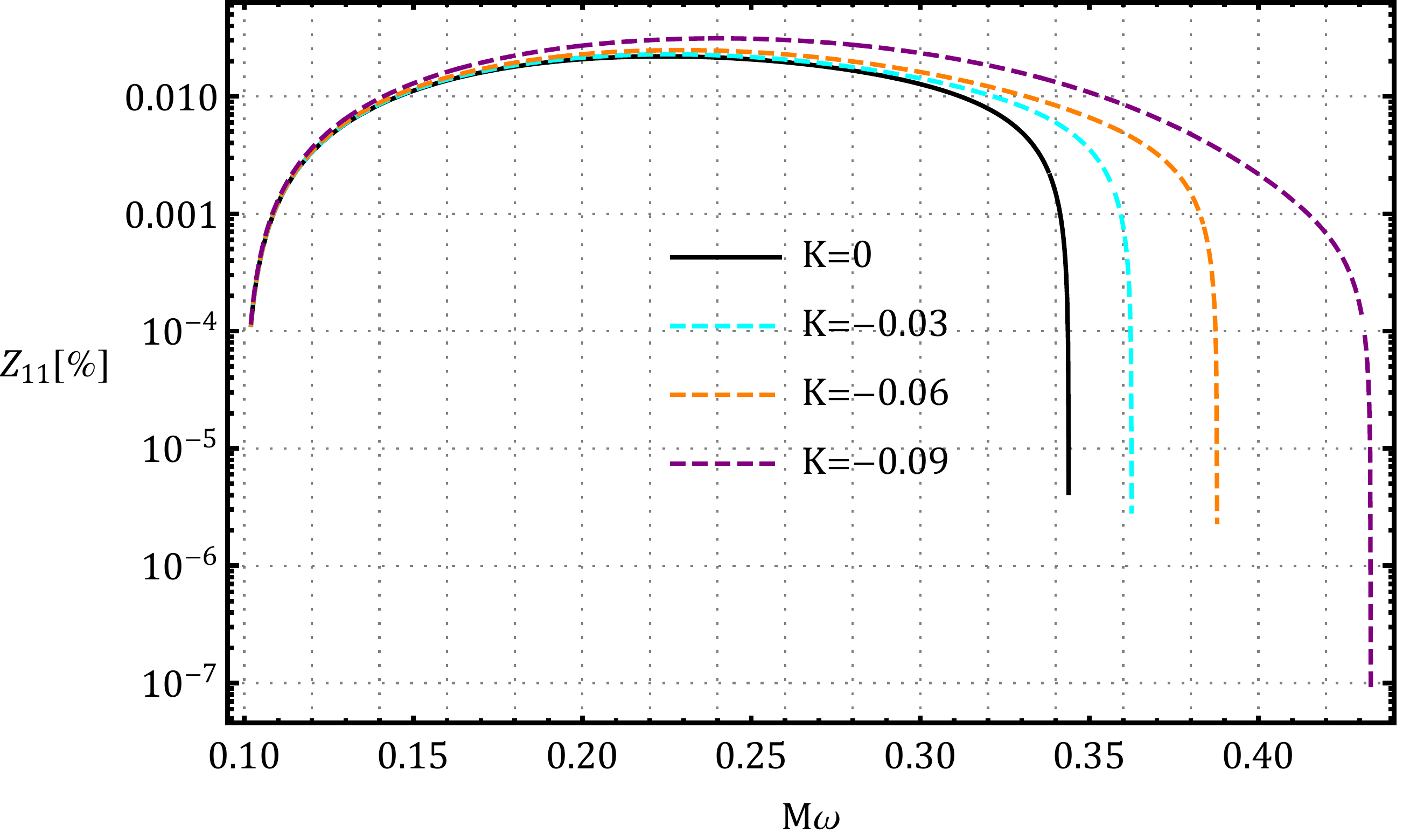}~~
		\includegraphics[scale=0.33]{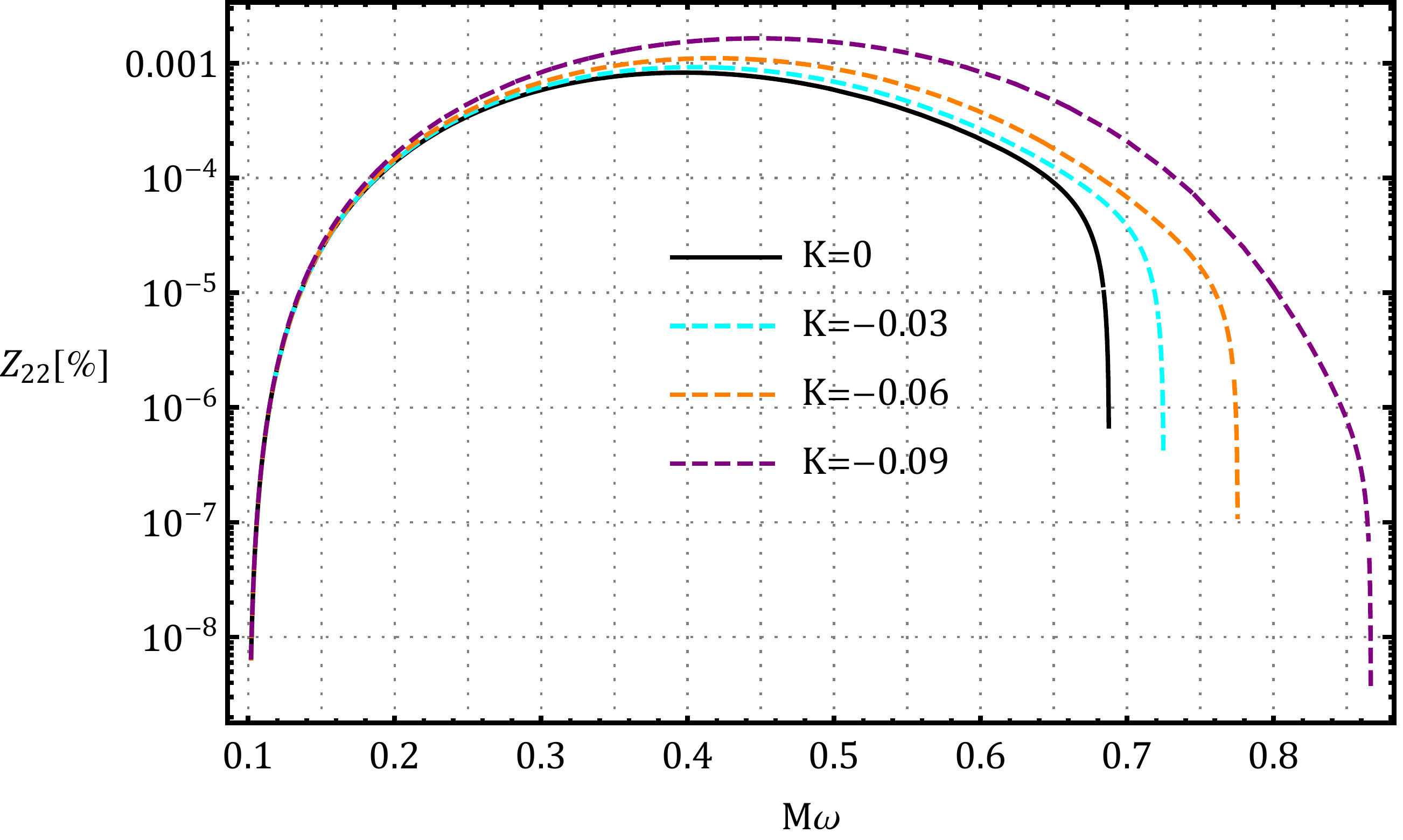}
		\includegraphics[scale=0.33]{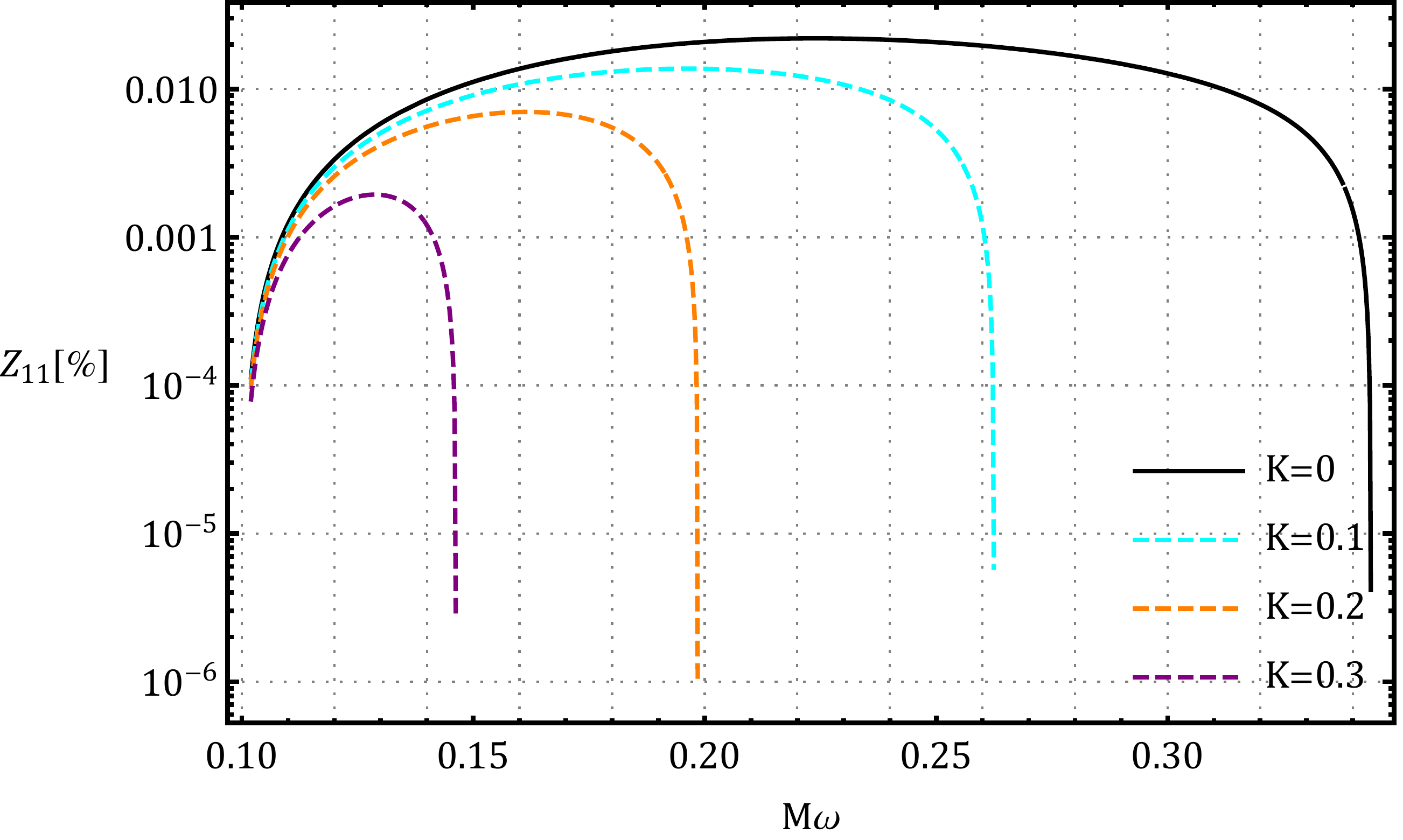}~~
		\includegraphics[scale=0.33]{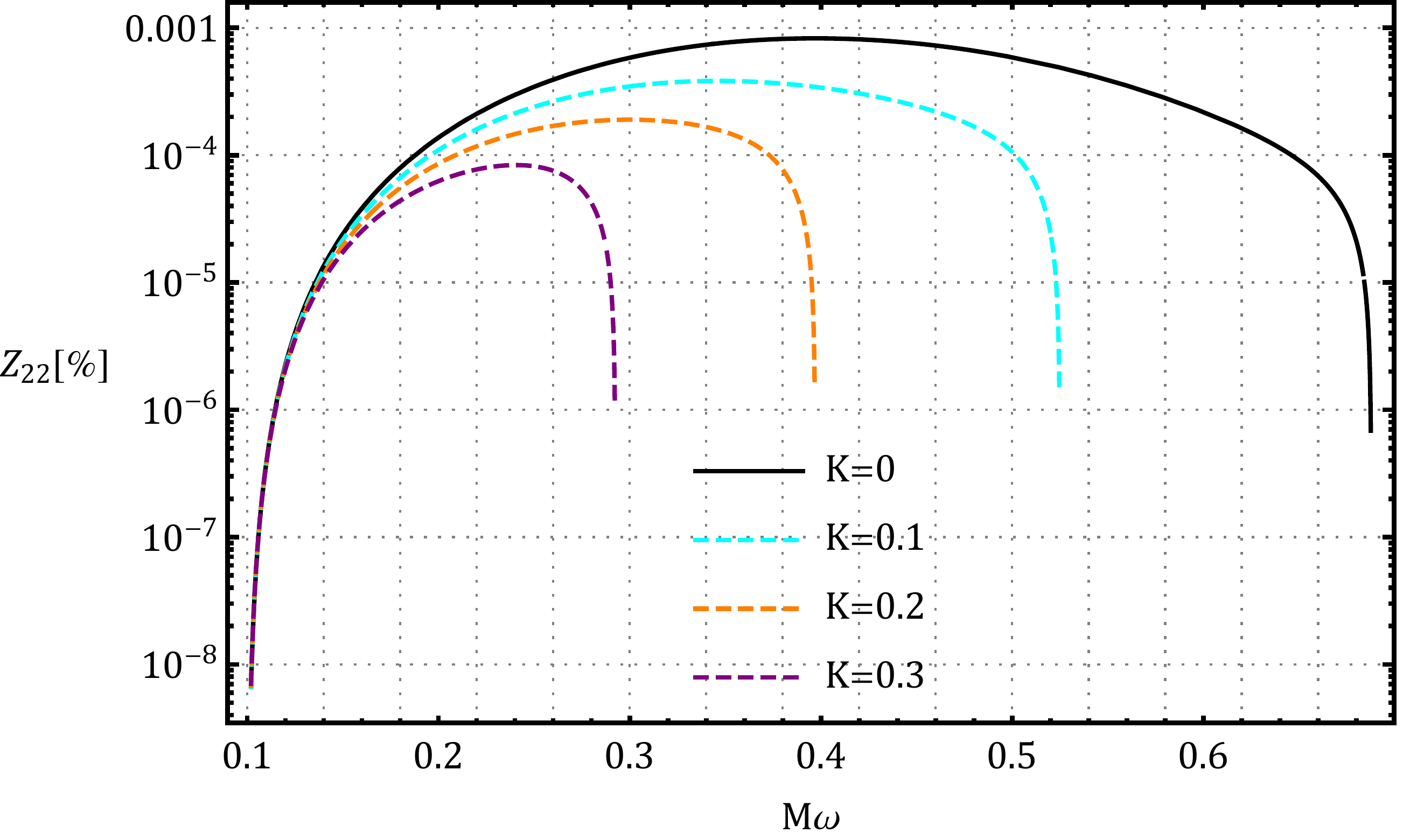}
		\caption{Plots $Z_{11}-M\omega$ and $Z_{22}-M\omega$ for the fluid matters: radiation ($\alpha=1/3$), and dark matter ($\alpha=-1/3$) in top and bottom rows, respectively. We use numerical value $0.95$ for the ratio of angular momentum to BH mass $a/M$.}
		\label{Zrdm}
	\end{center}
\end{figure}
\begin{figure}[!ht]
	\begin{center}
		\includegraphics[scale=0.33]{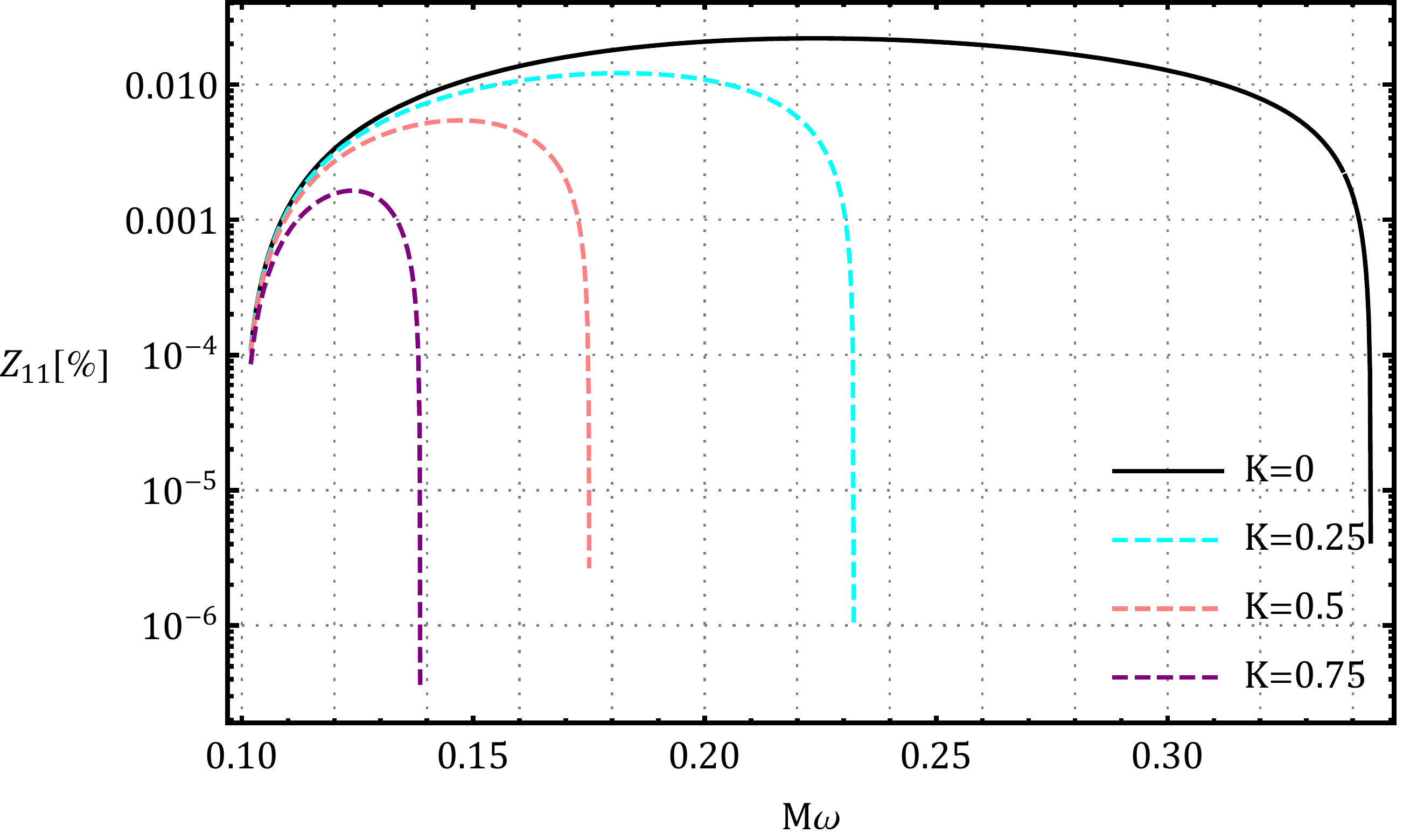}~~
		\includegraphics[scale=0.33]{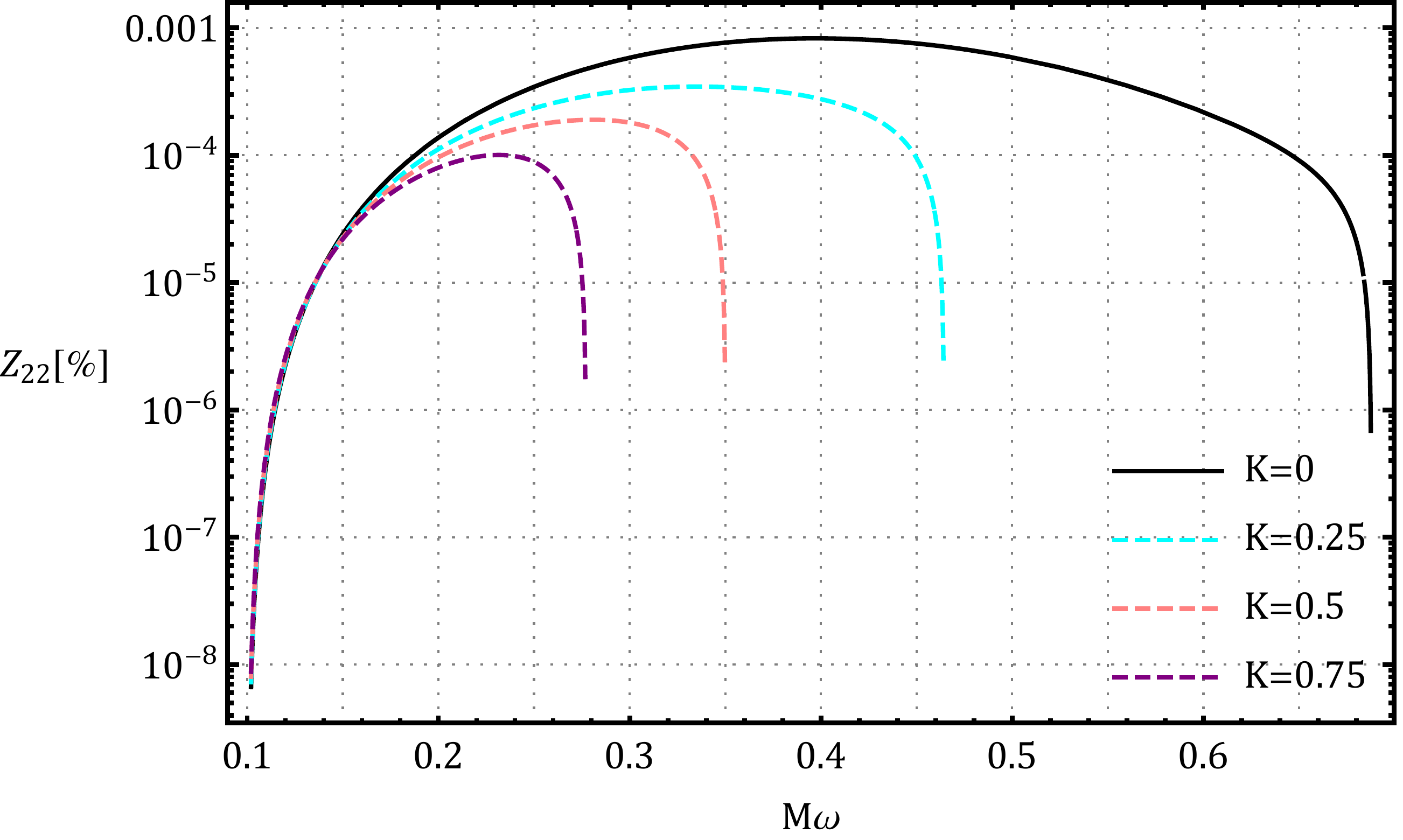}
		\includegraphics[scale=0.33]{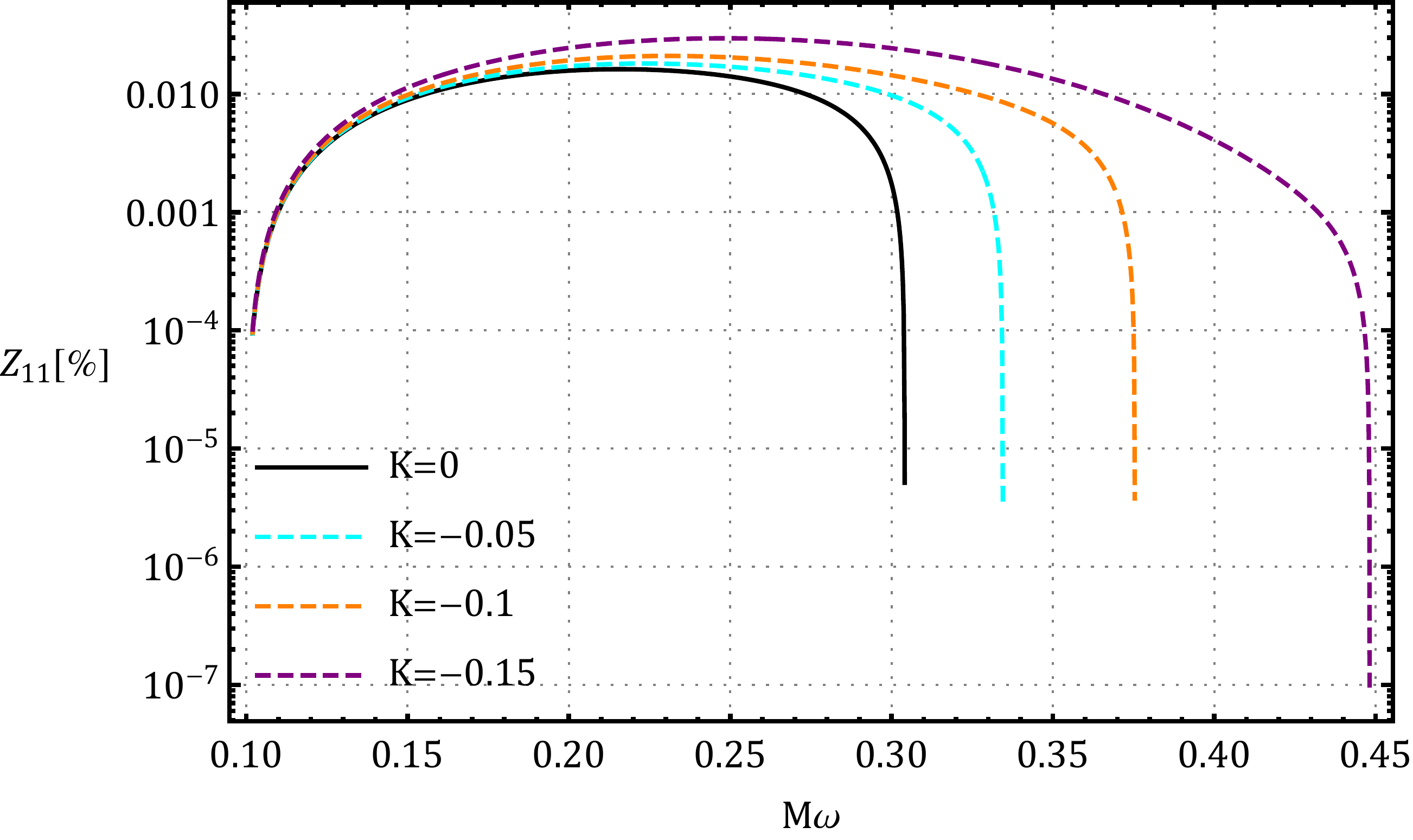}~~
		\includegraphics[scale=0.33]{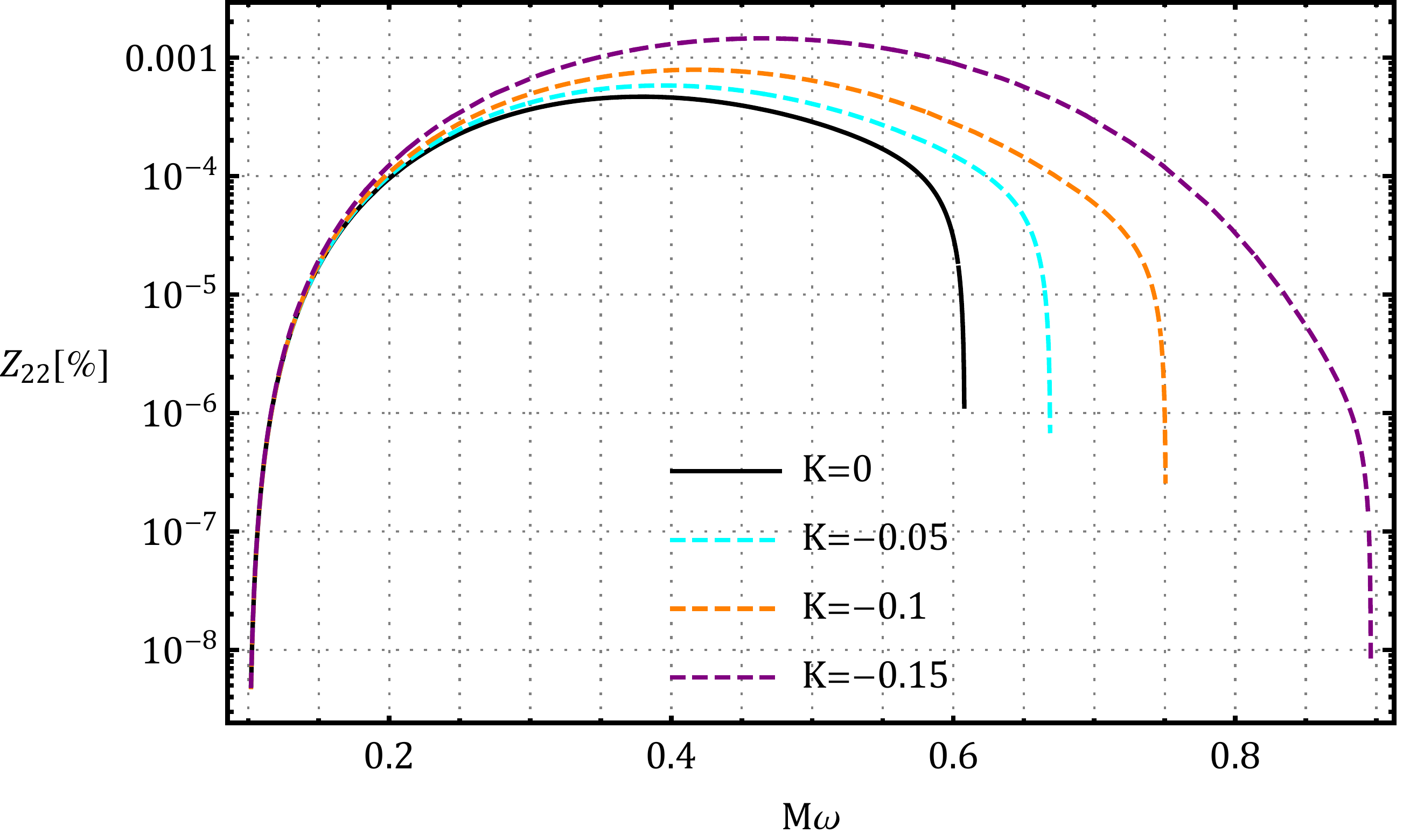}
		\caption{Plots $Z_{11}-M\omega$ and $Z_{22}-M\omega$ for the dust fluid matter ($\alpha=0$), for the positive and negative values of $K$ in top  and bottom rows, respectively. We use numerical value $0.95$ for the ratio of angular momentum to BH mass $a/M$.}
		\label{Zd}
	\end{center}
\end{figure}
\subsection{Outputs in the presence of anisotropic fluid matter}
Here, using \eqref{amp}, we release the results relevant to the leading multipoles ($l=1, 2$) of the massive scalar wave scattered off the Kiselev rotating BH with the anisotropic fluid matters: radiation ($\alpha=1/3$), dust ($\alpha=0$) and dark matter ($\alpha=-1/3$).
Indeed, $Z_{lm}$ is a gain factor assigned to the scalar wave mode 
which if $Z_{lm}>0$,  the superradiance happens, while $Z_{lm}<0$
denotes a loss factor meaning the lack of superradiance.
In Figs. (\ref{Zrdm}), and (\ref{Zd}) for every three types of matter fields, we draw the amplification factors $Z_{11}$ and $Z_{22}$ \footnote{As usual, the wave frequency $\omega$ is defined to be positive, and thereby, it is clear the modes with azimuthal wave numbers $m\leq0$ are not able to satisfy the superradiance condition $\omega<m\Omega_+$. Therefore, we are not of interest to these modes.}. These two figures give us two different messages. They obviously reveal to us if the Kerr BH enclose by radiation and dust (with $K<0$) matter fields, the scalar amplification factor, as well as the frequency range of superradiance, gets bigger and wider compared to $K=0$, respectively. In  other words, the presence of a profile of radiation and dust (with $K<0$) around a Kerr BH results in enhanced superradiance. However, in the case of enclosing the Kerr BH by a profile of dark matter and dust with $K>0$,  the superradiance is weakened. This is because the scalar amplification factor and the frequency range of superradiance get smaller and narrower compared to $K=0$, respectively. Concerning the case of radiation, the obtained output can also be utilized for the Kerr-Newman background which is subjected to the charged massive scalar perturbation. 
Although , for two reasons, the Kiselev metric enclosed by radiation seems more realistic compared to Kerr-Newman. First, it is well-known that the astrophysical BHs are electrically neutral. Second, the only massive scalar boson admitted by the standard model, Higgs, has no electric charge. The results derived for the case of dust indicate the effect of different values of the BH mass on the amplification factor \cite{Dolan:2007mj,Dolan:2012yt} since the Kiselev metric enclosed by dust is equivalent to the Kerr BH with a shifted mass. As mentioned already, the Kiselev rotating metric including dark matter fluid is similar to a Kerr BH enclosed by a cloud of cosmic strings. So it can be said, these two elements of matter leave the same effect on the massive scalar superradiance scattering, i.e., both weaken it.
 
\section{The effect of Anisotropic Fluid Matter on Stability Regime}\label{sec:stability}

In the previous section, we investigated the superradiant amplification for a scalar wave, scattered off from the Kerr BH surrounded by the anisotropic fluid matters (as addressed in the metric (\ref{Rot})).
In what follows, we want to explore the role of the underlying anisotropic fluid matters on superradiant stability through a mechanism  known as  \textit{``BH bomb''} \cite{Press:1972zz}. Speaking technically, capturing the massive modes of a system composed of the Kerr background (\ref{Rot}) and massive scalar perturbations $\Phi$, inside the effective potential well located outside the BH, may result in make instability in this system. Actually, these massive modes, behave similarly to a reflecting surface like the mirror, so that by returning the reflecting waves toward BH and their amplification and resonance via forward and backward moves, it gives rise to a superradiant instability known as the BH bomb. In addition to the existence of the ergo-region, as an essential component in the superradiant amplification, to trigger the instability a potential well outside the BH is required, too \cite{Hod:2012zza}. 
Beginning from the radial KG equation \eqref{eq:ODE_rad}, we have
\begin{equation}\label{eq:kg}
	\Delta_{k}{{d}\over{dr}}\Big(\Delta_{k}{{dF_{\omega lm}}\over{dr}}\Big)+\mathcal{U} F_{\omega lm}=0~,
\end{equation}
where
\begin{equation}\label{eq:M}
	\mathcal{U}\equiv \bigg((r^2+a^2)\omega-ma\bigg)^2+\Delta_k\bigg(2ma\omega-a^2\omega^2-l(l+1)-\mu_s^2r^2\bigg)~.
\end{equation}
To have the BH bomb, we must demand the following asymptotic solutions for the radial KG equation (\ref{eq:kg})
\begin{eqnarray}\label{eq:so}
	F_{\omega lm}\sim\left\{
	\begin{array}{ll}
		e^{-i (\omega-m\Omega_+)r_*}\ \ \text{ as }\ r\rightarrow r_{+}\ \
		(r_*\rightarrow -\infty)  \\\\
		\frac{e^{-\sqrt{\mu_s^2-\omega^2}r_*}}{r}\ \ \text{ as }\
		r\rightarrow\infty\ \ \ \ (r_*\rightarrow \infty)
	\end{array}
	\right.
\end{eqnarray} where the scalar wave on the BH horizon is purely ingoing; while, at spatial infinity it is a bounded solution i.e. decaying exponentially, if $\omega^2<\mu_s^2$. Inserting a new radial function as below
\begin{equation}\label{eq:n}
	\phi_{\omega lm}\equiv \sqrt{\Delta_k}F_{\omega lm}\  ,
\end{equation}
into the radial equation (\ref{eq:kg}), after some algebra we yield
the following Regge-Wheeler equation
\begin{equation}\label{eq:RW}
	\bigg({{d^2}\over{dr^2}}+\omega^2-V\bigg)\phi_{\omega lm}=0\  ,
\end{equation}
with
\begin{equation}\label{eq:RW2}
	V=\omega^2-{{f+\mathcal{U}}\over{\Delta_k^2}}\,,
\end{equation} where
\begin{equation}\label{}
	f=M^2-a^2+\frac{K^2(1-9\alpha^2)}{4}r^{-6\alpha}+\frac{K^2}{2}
	r^{-1-3\alpha}\bigg((9\alpha^2+3\alpha)r^2+M(2-18\alpha^2)r+(9\alpha^2-3\alpha)a^2~\bigg)~.
\end{equation} It is clear by relaxing the parameter $K$ in the above expression, the equation (\ref{eq:RW2}) comes back to its standard form. Now, by a straightforward computation, we can show that the asymptotic form of the effective potential $V$ (by ignoring the terms $\mathcal{O}(1/r^2)$), for the fluid matters: radiation ($\alpha=1/3$), dust ($\alpha=0$) and dark matter ($\alpha=-1/3$), can be written in the following form 
\begin{align}
	&V_{radiation}(r)=\mu_s^2-\frac{4M\omega^2-2M\mu_s^2}{r}~, \label{eq:Vr}\\
	&V_{dust}(r)=\mu_s^2-\frac{(4M+2K)\omega^2-(2M+2K)\mu_s^2}{r}~,\label{eq:Vd}\\
	&V_{dark~  matter}(r)=\frac{\mu_s^2 +(K^2-2K)\omega^2}{(K-1)^2}-\frac{(2K+2)M\mu_s^2-4M\omega^2}{(K-1)^3r}\textcolor{red}{}~, \label{eq:Vdm}
\end{align}
respectively. If the asymptotic derivative of the effective potential is positive, i.e., $V'\to 0^+$ as $r\to\infty$, it means the potential represents trapping well \cite{Hod:2012zza}. By demanding  that for the effective potentials (\ref{eq:Vr}) and (\ref{eq:Vd}), one can
acquire the following instability regime
\begin{equation}\label{eq:reg}
\frac{\mu_s^2}{2}<\omega^2<\mu_s^2\ .
\end{equation}
It essentially is nothing but the same regime in which the bound states of a system composed of the massive scalar field and the standard Kerr BH may become captured and result in instability in the background.
However we are interested in the superradiant instability.  Hence, we should merge the superradiant condition $\omega<m\Omega_+$, with instability regime (\ref{eq:reg}). This results in the superradiant instability regime $\frac{\mu_s}{\sqrt{2}}<\omega<m\Omega_+$, which 
in its complementary regime
\begin{equation}\label{eq:st}
\mu_s\geq\sqrt{2}m\Omega_+\ ,
\end{equation} the system composed of the massive scalar field and Kiselev rotating BH surrounded by the anisotropic fluid matters: radiation and dust, remains stable. As a consequence,  up to leading order, $\mathcal{O}(1/r^2)$, these two fluid matters are not able to affect the standard stability regime. Although, it is not true for the case with the fluid dark matter. By demanding $V'\to 0^+$ as $r\to\infty$ for the effective potential (\ref{eq:Vdm}) we have 
\begin{equation}\label{eq:regd}
	\frac{\mu_s^2}{2}(K+1)<\omega^2<\mu_s^2,~~~~0\leq K<1,
\end{equation} which finally result in the following superradiant stability regime
\begin{equation}\label{eq:std}
	\mu_s\geq\sqrt{\dfrac{2}{K+1}}m\Omega_+~.
\end{equation} It means that the presence of fluid dark matter around the Kerr BH reduces the lower bound of the scalar field mass required to ensure the stability of massive KG equation in the standard Kerr spacetime \cite{Hod:2012zza}. So, a rotating BH surrounded by the fluid dark matter, which is subject to the superradiant scattering of scalar perturbation with lower mass, has the chance of remain stable yet. All in all, by narrowing the instability regime (\ref{eq:regd}), thereby, the fluid dark matter can be thought of as an environmental component in the favor of BH stability.
\section{Conclusion}\label{sec:DC}

In this paper, we have investigated the role of the anisotropic fluid matter around a rotating BH on the amplification factor of the superradiance. Particularly, it has been done by employing the massive scalar perturbation scattered off the rotating Kiselev BH enclosed by three types of anisotropic fluid matters; radiation, dust, and dark matter. The corresponding spacetime metric contains two additional parameters, $\alpha$, and $K$, representing the EoS parameter and the intensity of fluid matter, respectively.
To guarantee the positive energy density condition, it is required that $\alpha K\leq0$. By considering that, for the radiation ($\alpha=1/3$) and the dark matter ($\alpha=-1/3$); respectively, we have $K\leq0$, and $K\geq0$, while; for the dust ($\alpha=0$), both cases are possible.
By conducting our analysis within the low-frequency and low-mass regimes
for the scalar perturbation, we have shown the amplification factor as well as its frequency range, both are affected by the parameter $K$. More precisely, we found $K<0$, and $K>0$, enhances and suppresses the massive scalar superradiance scattering compared to the case $K=0$, respectively. The frequency range of superradiance also turns broader and narrower than $K=0$, respectively. As a result, the presence of radiation and dark matter around a rotating BH can play a role similar to the amplifiers and attenuators for superradiance, respectively. While the dust due to admitting both positive and negative signs for the parameter $K$ can play a twofold role.

At the end, by taking the BH bomb mechanism into the viewpoint of the effective potential, we have studied the superradiant stability concerning the rotating Kiselev BH, subjected by the massive scalar perturbation.
Through the effective potential analysis, we found only the dark matter has an effect on the standard instability regime $\frac{\mu_s^2}{2}<\omega^2<\mu_s^2$, so that by narrowing it as  a form of  $\frac{\mu_s^2}{2}(K+1)<\omega^2<\mu_s^2$, it increases the stability chance of BH against the massive scalar wave perturbation. It results in the reduction of the lower bound of the scalar field mass required to ensure the superradiant stability concerning the massive KG equation in the standard Kerr spacetime, i.e., $\mu_s\geq\sqrt{\dfrac{2}{K+1}}m\Omega_+$, where $0\leq K<1$. To sum it up, this result indicates the fluid dark matter around Kerr BH can play an effective role in favor of BH superradiant stability. 

\section*{Acknowledgements}
M. Kh would like to thank Carlos Herdeiro for his thorough reading and enlightening comments on the manuscript. M. Kh is also grateful to Kimet Jusufi for the useful discussions and suggestions. R. P would like to express gratitude to Hamid Reza Sepangi for all his encouragement and support.



\begin{thebibliography}{99}
	
\bibitem{Akiyama:2019fyp}
K.~Akiyama {\it et al.} [Event Horizon Telescope Collaboration],
\textcolor{cyan}{Astrophys.\ J.\ Lett.\  {\bf 875} (2019) no.1, L5
[arXiv:1906.11242 [astro-ph.GA]].}


\bibitem{Comisso:2020ykg}
L.~Comisso and F.~A.~Asenjo,
\textcolor{cyan}{Phys.\ Rev.\ D {\bf 103} (2021) no.2,  023014
[arXiv:2012.00879 [astro-ph.HE]].}


\bibitem{Penrose:1969pc}
R.~Penrose,
\textcolor{cyan}{Riv.\ Nuovo Cim.\  {\bf 1} (1969) 252
[Gen.\ Rel.\ Grav.\  {\bf 34} (2002) 1141].}

\bibitem{Zel:1971}
Ya. B. Zel'dovich, \textcolor{cyan}{JETP Lett. \textbf{14} (1971) 180.}

\bibitem{Zel:1972}
Ya. B. Zel'dovich, \textcolor{cyan}{Sov. Phys. JETP \textbf{35}  (1972) 1085.}

\bibitem{Starobinsky:1973aij}
A.~A.~Starobinsky,
\textcolor{cyan}{Sov. Phys. JETP \textbf{37} (1973) no.1, 28-32.}

\bibitem{TP:1974}
S.A. Teukolsky, W.H. Press, \textcolor{cyan}{Astrophys.J. 193 (1974) 443-461.}


\bibitem{Bekenstein:1998nt}
J.~D.~Bekenstein and M.~Schiffer,
\textcolor{cyan}{Phys. Rev. D \textbf{58} (1998), 064014
[arXiv:gr-qc/9803033 [gr-qc]].}

\bibitem{Brito:2015oca}
R.~Brito, V.~Cardoso and P.~Pani,
\textcolor{cyan}{Lect. Notes Phys. \textbf{906} (2015), pp.1-237
[arXiv:1501.06570 [gr-qc]].}


\bibitem{Bekenstein:1973mi}
J.~D.~Bekenstein,
\textcolor{cyan}{Phys. Rev. D \textbf{7} (1973), 949-953.}

\bibitem{DiMenza:2014vpa}
L.~Di Menza and J.~P.~Nicolas,
\textcolor{cyan}{Class. Quant. Grav. \textbf{32} (2015) no.14, 145013
[arXiv:1411.3988 [math-ph]].}


\bibitem{Detweiler:1980uk}
S.~L.~Detweiler,
\textcolor{cyan}{Phys.\ Rev.\ D {\bf 22}, 2323 (1980).}


\bibitem{Hod:2016bas}
S.~Hod,
\textcolor{cyan}{Phys. Rev. D \textbf{94} (2016) no.4, 044036
[arXiv:1609.07146 [gr-qc]].}

\bibitem{Benone:2019all}
C.~L.~Benone and L.~C.~B.~Crispino,
\textcolor{cyan}{Phys. Rev. D \textbf{99} (2019) no.4, 044009
[arXiv:1901.05592 [gr-qc]].}


\bibitem{Pani:2011gy}
P.~Pani, C.~F.~B.~Macedo, L.~C.~B.~Crispino and V.~Cardoso,
\textcolor{cyan}{Phys.\ Rev.\ D {\bf 84} (2011) 087501
[arXiv:1109.3996 [gr-qc]].}

\bibitem{Kleihaus:2011tg}
B.~Kleihaus, J.~Kunz and E.~Radu,
\textcolor{cyan}{Phys.\ Rev.\ Lett.\  {\bf 106} (2011) 151104
[arXiv:1101.2868 [gr-qc]].}

\bibitem{Delsate:2018ome}
T.~Delsate, C.~Herdeiro and E.~Radu,
\textcolor{cyan}{Phys.\ Lett.\ B {\bf 787} (2018) 8
[arXiv:1806.06700 [gr-qc]].}

\bibitem{Cunha:2019dwb}
P.~V.~P.~Cunha, C.~A.~R.~Herdeiro and E.~Radu,
\textcolor{cyan}{Phys.\ Rev.\ Lett.\  {\bf 123} (2019) no.1,  011101
[arXiv:1904.09997 [gr-qc]].}


\bibitem{Cardoso:2013opa}
V.~Cardoso, I.~P.~Carucci, P.~Pani and T.~P.~Sotiriou,
\textcolor{cyan}{Phys.\ Rev.\ D {\bf 88} (2013) 044056
[arXiv:1305.6936 [gr-qc]].}

\bibitem{Cardoso:2013fwa}
V.~Cardoso, I.~P.~Carucci, P.~Pani and T.~P.~Sotiriou,
\textcolor{cyan}{Phys.\ Rev.\ Lett.\  {\bf 111} (2013) 111101
[arXiv:1308.6587 [gr-qc]].}

\bibitem{Aliev:2014aba}
A.~N.~Aliev,
\textcolor{cyan}{JCAP {\bf 1411} (2014) 029
[arXiv:1408.4269 [hep-th]].}

\bibitem{Zhang:2014kna}
C.~Y.~Zhang, S.~J.~Zhang and B.~Wang,
\textcolor{cyan}{JHEP {\bf 1408} (2014) 011
[arXiv:1405.3811 [hep-th]].}

\bibitem{Fierro:2017fky}
O.~Fierro, N.~Grandi and J.~Oliva,
\textcolor{cyan}{Class.\ Quant.\ Grav.\  {\bf 35} (2018) no.10,  105007
[arXiv:1708.06037 [hep-th]].}

\bibitem{Wondrak:2018fza}
M.~F.~Wondrak, P.~Nicolini and J.~W.~Moffat,
\textcolor{cyan}{JCAP {\bf 1812} (2018) 021
[arXiv:1809.07509 [gr-qc]].}

\bibitem{Kolyvaris:2018zxl}
T.~Kolyvaris, M.~Koukouvaou, A.~Machattou and E.~Papantonopoulos,
\textcolor{cyan}{Phys.\ Rev.\ D {\bf 98} (2018) no.2,  024045
[arXiv:1806.11110 [gr-qc]].}

\bibitem{Rahmani:2020wlq}
A.~Rahmani, M.~Honardoost and H.~R.~Sepangi,
\textcolor{cyan}{Phys.\ Rev.\ D {\bf 101} (2020) no.8,  084036
[arXiv:2002.01663 [gr-qc]].}

\bibitem{Khodadi:2020cht}
M.~Khodadi, A.~Talebian and H.~Firouzjahi,
\textcolor{cyan}{arXiv:2002.10496 [gr-qc].}

\bibitem{Zhang:2020sjh}
C.~Y.~Zhang, S.~J.~Zhang, P.~C.~Li and M.~Guo,
\textcolor{cyan}{JHEP \textbf{08} (2020), 105
[arXiv:2004.03141 [gr-qc]].}


\bibitem{Khodadi:2021owg}
M.~Khodadi,
\textcolor{cyan}{Phys. Rev. D \textbf{103} (2021) no.6, 064051
[arXiv:2103.03611 [gr-qc]].}

\bibitem{Mehta:2021pwf}
V.~M.~Mehta, M.~Demirtas, C.~Long, D.~J.~E.~Marsh, L.~McAllister and M.~J.~Stott,
\textcolor{cyan}{JCAP \textbf{07} (2021), 033
[arXiv:2103.06812 [hep-th]].}

\bibitem{Jiang:2021whw}
R.~Jiang, R.~H.~Lin and X.~H.~Zhai,
\textcolor{cyan}{[arXiv:2108.04702 [gr-qc]].}

\bibitem{Witek:2012tr}
H.~Witek, V.~Cardoso, A.~Ishibashi and U.~Sperhake,
\textcolor{cyan}{Phys. Rev. D \textbf{87} (2013) no.4, 043513
[arXiv:1212.0551 [gr-qc]].}


\bibitem{Cardoso:2004nk}
V.~Cardoso, O.~J.~C.~Dias, J.~P.~S.~Lemos and S.~Yoshida,
\textcolor{cyan}{Phys. Rev. D \textbf{70} (2004), 044039
[erratum: Phys. Rev. D \textbf{70} (2004), 049903]
[arXiv:hep-th/0404096 [hep-th]].}


\bibitem{Cardoso:2013krh}
V.~Cardoso,
\textcolor{cyan}{Gen. Rel. Grav. \textbf{45} (2013), 2079-2097
[arXiv:1307.0038 [gr-qc]].}

\bibitem{Degollado:2013bha}
J.~C.~Degollado and C.~A.~R.~Herdeiro,
\textcolor{cyan}{Phys.\ Rev.\ D {\bf 89} (2014) no.6,  063005
[arXiv:1312.4579 [gr-qc]].}

\bibitem{Herdeiro:2013pia}
C.~A.~R.~Herdeiro, J.~C.~Degollado and H.~F.~Rúnarsson,
\textcolor{cyan}{Phys.\ Rev.\ D {\bf 88} (2013) 063003
[arXiv:1305.5513 [gr-qc]].}

\bibitem{Dolan:2015dha}
S.~R.~Dolan, S.~Ponglertsakul and E.~Winstanley,
\textcolor{cyan}{Phys.\ Rev.\ D {\bf 92} (2015) no.12,  124047
[arXiv:1507.02156 [gr-qc]].}

\bibitem{Dias:2018zjg}
O.~J.~C.~Dias and R.~Masachs,
\textcolor{cyan}{Class.\ Quant.\ Grav.\  {\bf 35} (2018) no.18,  184001
[arXiv:1801.10176 [gr-qc]].}


\bibitem{Press:1972zz}
W.~H.~Press and S.~A.~Teukolsky,
\textcolor{cyan}{Nature \textbf{238} (1972), 211-212.}


\bibitem{Hod:2014pza}
S.~Hod,
\textcolor{cyan}{Phys.\ Lett.\ B {\bf 736} (2014) 398
[arXiv:1412.6108 [gr-qc]].}

\bibitem{Hod:2016rqd}
S.~Hod,
\textcolor{cyan}{Phys.\ Lett.\ B {\bf 761} (2016) 326
[arXiv:1612.02819 [gr-qc]].}


\bibitem{Furuhashi:2004jk}
H.~Furuhashi and Y.~Nambu,
\textcolor{cyan}{Prog. Theor. Phys. \textbf{112} (2004), 983-995
[arXiv:gr-qc/0402037 [gr-qc]].}


\bibitem{Cardoso:2006wa}
V.~Cardoso, O.~J.~C.~Dias and S.~Yoshida,
\textcolor{cyan}{Phys.\ Rev.\ D {\bf 74} (2006) 044008
[hep-th/0607162].}


\bibitem{Dolan:2007mj}
S.~R.~Dolan,
\textcolor{cyan}{Phys. Rev. D \textbf{76} (2007), 084001
[arXiv:0705.2880 [gr-qc]].}

\bibitem{Hod:2012zza}
S.~Hod,
\textcolor{cyan}{Phys.\ Lett.\ B {\bf 708} (2012) 320
[arXiv:1205.1872 [gr-qc]].}


\bibitem{Li:2012rx}
R.~Li,
\textcolor{cyan}{Phys.\ Lett.\ B {\bf 714} (2012) 337
[arXiv:1205.3929 [gr-qc]].}

\bibitem{Dolan:2012yt} 
S.~R.~Dolan,
\textcolor{cyan}{Phys. Rev. D \textbf{87} (2013) no.12, 124026
[arXiv:1212.1477 [gr-qc]].}


\bibitem{Zhu:2014sya}
Z.~Zhu, S.~J.~Zhang, C.~E.~Pellicer, B.~Wang and E.~Abdalla,
\textcolor{cyan}{Phys. Rev. D \textbf{90} (2014) no.4, 044042
[arXiv:1405.4931 [hep-th]].}


\bibitem{Green:2015kur}
S.~R.~Green, S.~Hollands, A.~Ishibashi and R.~M.~Wald,
\textcolor{cyan}{Class.\ Quant.\ Grav.\  {\bf 33} (2016) no.12,  125022
[arXiv:1512.02644 [gr-qc]].}


\bibitem{Huang:2018qdl}
Y.~Huang, D.~J.~Liu, X.~h.~Zhai and X.~z.~Li,
\textcolor{cyan}{Phys. Rev. D \textbf{98} (2018) no.2, 025021
[arXiv:1807.06263 [gr-qc]].}


\bibitem{Destounis:2019hca}
K.~Destounis,
\textcolor{cyan}{Phys.\ Rev.\ D {\bf 100} (2019) no.4,  044054
[arXiv:1908.06117 [gr-qc]].}

\bibitem{Huang:2019xbu}
J.~H.~Huang, W.~X.~Chen, Z.~Y.~Huang and Z.~F.~Mai,
\textcolor{cyan}{Phys.\ Lett.\ B {\bf 798} (2019) 135026
[arXiv:1907.09118 [gr-qc]].}

\bibitem{Li:2019tns}
R.~Li, Y.~Zhao, T.~Zi and X.~Chen,
\textcolor{cyan}{Phys.\ Rev.\ D {\bf 99} (2019) no.8,  084045.}

\bibitem{Xu:2020fgq}
J.~H.~Xu, Z.~H.~Zheng, M.~J.~Luo and J.~H.~Huang,
\textcolor{cyan}{Eur. Phys. J. C \textbf{81} (2021) no.5, 402
[arXiv:2012.13594 [gr-qc]].}

\bibitem{Vieira:2021nha}
H.~S.~Vieira, V.~B.~Bezerra and C.~R.~Muniz,
\textcolor{cyan}{[arXiv:2107.02562 [gr-qc]].}


\bibitem{Herdeiro:2016tmi}
C.~Herdeiro, E.~Radu and H.~Rúnarsson,
\textcolor{cyan}{Class.\ Quant.\ Grav.\  {\bf 33} (2016) no.15,  154001
[arXiv:1603.02687 [gr-qc]].}

\bibitem{Herdeiro:2017phl}
C.~A.~R.~Herdeiro and E.~Radu,
\textcolor{cyan}{Phys.\ Rev.\ Lett.\  {\bf 119} (2017) no.26,  261101
[arXiv:1706.06597 [gr-qc]].}

\bibitem{Degollado:2018ypf}
J.~C.~Degollado, C.~A.~R.~Herdeiro and E.~Radu,
\textcolor{cyan}{Phys. Lett. B \textbf{781} (2018), 651-655
[arXiv:1802.07266 [gr-qc]].}

\bibitem{Herdeiro:2020xmb}
C.~A.~R.~Herdeiro and E.~Radu,
\textcolor{cyan}{Eur.\ Phys.\ J.\ C {\bf 80} (2020) no.5,  390
[arXiv:2004.00336 [gr-qc]].}


\bibitem{Rahmani:2020vvv}
A.~Rahmani, M.~Khodadi, M.~Honardoost and H.~R.~Sepangi,
\textcolor{cyan}{Nucl.\ Phys.\ B {\bf 960} (2020) 115185
[arXiv:2009.09186 [gr-qc]].}


\bibitem{Brito:2014wla}
R.~Brito, V.~Cardoso and P.~Pani,
\textcolor{cyan}{Class. Quant. Grav. \textbf{32} (2015) no.13, 134001
[arXiv:1411.0686 [gr-qc]].}


\bibitem{Aliev:2008yk}
A.~N.~Aliev and O.~Delice,
\textcolor{cyan}{Phys. Rev. D \textbf{79} (2009), 024013
[arXiv:0808.0280 [hep-th]].}


\bibitem{Wang:2014eha}
M.~Wang and C.~Herdeiro,
\textcolor{cyan}{Phys. Rev. D \textbf{89} (2014) no.8, 084062
[arXiv:1403.5160 [gr-qc]].}

\bibitem{Huang:2016zoz}
Y.~Huang, D.~J.~Liu and X.~Z.~Li,
\textcolor{cyan}{Int. J. Mod. Phys. D \textbf{26} (2017) no.13, 1750141
[arXiv:1606.00100 [gr-qc]].}

\bibitem{Ishibashi:2015rya}
A.~Ishibashi, P.~Pani, L.~Gualtieri and V.~Cardoso,
\textcolor{cyan}{JHEP \textbf{09} (2015), 209
[arXiv:1507.07079 [hep-th]].}



\bibitem{Kiselev:2002dx}
V.~V.~Kiselev,
\textcolor{cyan}{Class. Quant. Grav. \textbf{20} (2003), 1187-1198
[arXiv:gr-qc/0210040 [gr-qc]].}

\bibitem{Toshmatov:2015npp}
B.~Toshmatov, Z.~Stuchl\'\i{}k and B.~Ahmedov,
\textcolor{cyan}{Eur. Phys. J. Plus \textbf{132} (2017) no.2, 98
[arXiv:1512.01498 [gr-qc]].}


\bibitem{Cuadros-Melgar:2021sjy}
B.~Cuadros-Melgar, R.~D.~B.~Fontana and J.~de Oliveira,
\textcolor{cyan}{[arXiv:2108.04864 [gr-qc]].}

\bibitem{Cho:2017nhx}
I.~Cho and H.~C.~Kim,
\textcolor{cyan}{Chin. Phys. C \textbf{43} (2019) no.2, 025101
[arXiv:1703.01103 [gr-qc]].}

\bibitem{Kim:2019hfp}
H.~C.~Kim, B.~H.~Lee, W.~Lee and Y.~Lee,
\textcolor{cyan}{Phys. Rev. D \textbf{101} (2020) no.6, 064067
[arXiv:1912.09709 [gr-qc]].}


\bibitem{Hui:2016ltb}
L.~Hui, J.~P.~Ostriker, S.~Tremaine and E.~Witten,
\textcolor{cyan}{Phys. Rev. D \textbf{95} (2017) no.4, 043541
[arXiv:1610.08297 [astro-ph.CO]].}


\bibitem{Newman:1965tw}
E.~T.~Newman and A.~I.~Janis,
\textcolor{cyan}{J. Math. Phys. \textbf{6} (1965), 915-917.}

\bibitem{Azreg-Ainou:2014pra}
M.~Azreg-A\"\i{}nou,
\textcolor{cyan}{Phys. Rev. D \textbf{90} (2014) no.6, 064041
[arXiv:1405.2569 [gr-qc]].}
	
	
\bibitem{Rahaman:2010xs}
F.~Rahaman, K.~K.~Nandi, A.~Bhadra, M.~Kalam and K.~Chakraborty,
\textcolor{cyan}{Phys. Lett. B \textbf{694} (2011), 10-15
[arXiv:1009.3572 [gr-qc]].}
	
\bibitem{Xu:2018mkl}
Z.~Xu, X.~Hou and J.~Wang,
\textcolor{cyan}{JCAP \textbf{10} (2018), 046
[arXiv:1806.09415 [gr-qc]].}


\bibitem{Letelier:1979ej}
P.~S.~Letelier,
\textcolor{cyan}{Phys. Rev. D \textbf{20} (1979), 1294-1302}
	
\bibitem{Visser:2019brz}
M.~Visser,
\textcolor{cyan}{Class. Quant. Grav. \textbf{37} (2020) no.4, 045001
[arXiv:1908.11058 [gr-qc]].}
	

\end{thebibliography}
\end{document}